\documentclass[twocolumn,prb,showpacs,preprintnumbers,amsmath,amssymb,superscriptaddress,dvipdfmx]{revtex4}
\usepackage{graphicx,bm}
\usepackage{color}
\usepackage{setspace}

\renewcommand{\rceil}{\mbox{$\neg$}}
\renewcommand{\lceil}{\reflectbox{$\neg$}}

\begin{document}

\title{Interaction quench in the Holstein model: \\
Thermalization crossover from electron- to phonon-dominated relaxation}

\author{Yuta Murakami}
\affiliation{Department of Physics, The University of Tokyo, Hongo, Tokyo 113-0033, Japan}
\author{Philipp Werner}
\affiliation{Department of Physics, University of Fribourg, 1700 Fribourg, Switzerland}
\author{Naoto Tsuji}
\affiliation{Department of Physics, The University of Tokyo, Hongo, Tokyo 113-0033, Japan}
\author{Hideo Aoki}
\affiliation{Department of Physics, The University of Tokyo, Hongo, Tokyo 113-0033, Japan}

\date{\today}

\begin{abstract}
We study the relaxation of the Holstein model after a sudden switch-on of the interaction by means of the nonequilibrium 
dynamical mean field theory, 
with the self-consistent Migdal approximation as an impurity solver. We show that there exists a qualitative change in the thermalization dynamics as the interaction is varied in the weak-coupling regime. 
On the weaker interaction side of this crossover, the phonon oscillations are damped more rapidly than the electron thermalization timescale, as determined from the relaxation of the electron momentum distribution function. On the stronger interaction side, the relaxation of the electrons becomes faster than the phonon damping. 
In this regime, despite long-lived phonon oscillations, a thermalized momentum distribution is realized temporarily.  
The origin of the ``thermalization crossover" found here 
is traced back
to  different behaviors of the electron and phonon self-energies as a function of the electron-phonon coupling. 
In addition, the importance of the phonon dynamics is demonstrated  
by comparing the self-consistent Migdal results with those obtained with a simpler Hatree-Fock impurity solver that neglects the phonon self-energy. The latter scheme does not properly describe the evolution and thermalization of isolated electron-phonon systems.
\end{abstract}

\pacs{71.38.-k,71.10.Fd,71.10.-w}

\maketitle
\setstretch{0.9}
\section{\label{sec:intro}Introduction}
The nonequilibrium dynamics of correlated lattice systems has recently been investigated intensively in various contexts.\cite{Aoki2013} Interaction-quench studies\cite{Manmana2007, Kollath2007, MoeckelKehrein2008,Eckstein2009,Barmettler2009,Eckstein2010,Tsuji2013,Tsuji2013b,Hugo2014,Rigol2009,Rigol2009b,Rigol2011,Rigol2012,Rigol2013,Rigol2014} have been motivated by cold-atom experiments, where the interaction or hopping can be tuned by the Feshbach resonance or by changing the depth of the optical lattice potential.  
In condensed-matter experiments, on the other hand, one can drive correlated electron systems with strong lasers. These perturbations may induce phase transitions, e.g. from an insulating to a metallic state,\cite{Iwai2003,Perfetti2006,Okamoto2007,Okamoto2011} or in some cases metastable phases with interesting properties.\cite{Fausti2011, Mihailovic2014}
In real materials, the electron-phonon coupling can play a crucial role in the nonequilibrium dynamics, and indeed many pump-probe experiments exhibit clear signatures of phonon oscillations.\cite{Perfetti2006, Johnson2009} 
From a theoretical point of view, the interplay of electronic and lattice degrees of freedom in out-of-equilibrium situations 
is still far from fully understood. 
Conventionally, the experimental results are interpreted in terms of a phenomenological two-temperature model,\cite{Allen1987} 
which is based on the Boltzmann equation and the assumption that the electrons and phonons are in thermal %temporal 
equilibrium with respective, time-dependent temperatures. The Boltzmann equation itself has also been studied numerically\cite{Groeneveld1995} and analytically.\cite{Alexandrov2008} 

On the other hand, various techniques have been developed and used in recent years to study nonequilibrium electron-phonon systems on the microscopic level beyond the Boltzmann equation (gradient approxiamtion). 
In Refs.~\onlinecite{Vidmer2011,Denis2012,Denis2012b}, the non-equilibrium dynamics of one or two polarons in the Holstein model was investigated with a time-dependent exact diagonalization method.
As for many-electron problems, previous works have investigated systems with classical phonons\cite{Yonemitsu2009} or quantum phonons\cite{Matsueda2012} in one-dimension.
 Many-electron problems in two-dimensional systems have been studied with an exact diagonalization method for the Holstein-Hubbard model\cite{Filippis2012} or with a weak-coupling perturbation theory for the Holstein model.\cite{Sentef2013,Kemper2013,Kemper2014}  On the other hand, the dynamical mean-field theory (DMFT),\cite{Georges1996} which becomes exact in infinite spatial dimensions, has been used to study the interplay of electrons and phonons in the Mott insulating phase. These simulations, which employed a strong-coupling impurity solver,\cite{Werner2013,Werner2014} showed that the feedback of the lattice dynamics on the electrons can lead to significant changes in the spectral function, and to qualitatively different relaxation pathways. We also notice that this method has recently been applied to the single-electron problem in the non-equilibrium Holstein model\cite{Sayyad2014}.

Despite these advances, 
studies treating the dynamics of quantum phonons are so far mostly
limited to systems in or near the Mott insulating phase.\cite{Matsueda2012,Filippis2012,Werner2013,Werner2014} 
Hence it remains to be clarified how an electron-phonon system thermalizes in weakly or moderately correlated metallic systems and how the phonon dynamics affects the relaxation process beyond the conventional analysis based on the Boltzmann equation.\cite{Allen1987,Groeneveld1995,Alexandrov2008} 
In addition, various interesting questions that have been addressed in purely electronic systems (such as the Hubbard model) remain to be answered.
 For example, one may ask whether or not the so-called prethermalization phenomena\cite{Berges2004,MoeckelKehrein2008,Eckstein2009,Eckstein2010} and dynamical phase transitions\cite{Eckstein2009,Eckstein2010,Schiro2010,Schiro2011,Tsuji2013b} occur in electron-phonon systems. 
Obtaining insights into the effects of the phonon dynamics is also important for establishing suitable approximate treatments and their limitations.

To address the above issues, we focus in this paper on the simplest possible model for an electron-phonon system, i.e., the Holstein model, which contains the coupling to local (Einstein) phonons with neither Coulomb interactions nor a coupling to some phenomenological heat bath. 
We also consider the simplest possible kind of perturbation, i.e., we drive the system out of equilibrium by a sudden quench of the electron-phonon coupling. 
Such a quench
may be realized in cold-atom systems in optical lattices,\cite{Pazy2005,Herrera2011,Hague2012,Hohenadler2013}
and it is expected to be closely related to the phonon frequency quench, which has been studied experimentally in bismuth.\cite{Johnson2009}
Understanding the dynamics in this simple setup will provide a basis for the study of more complicated or realistic situations, e.g. photo-excitated systems, models with the Coulomb interaction or with acoustic phonons. 

The Holstein model is solved with the nonequilibrium extension of DMFT,\cite{Freericks2006,Aoki2013} which is exact in the limit of infinite spatial dimensions. Therefore, our results are relevant to systems with high spatial dimensions or large coordination numbers.
As an impurity solver for DMFT, we employ the self-consistent Migdal approximation.  
While this approximation neglects the vertex correction in the self-energy and is based on the assumption that the phonon frequency and the electron-phonon coupling are small, % 
this type of approximation has been successfully used to describe conventional superconductors in the 
correlated regime, and has also been justified with numerical studies based on the DMFT framework.\cite{Bauer2011}

\begin{figure}[t]
  \centering
   \includegraphics[width=70mm]{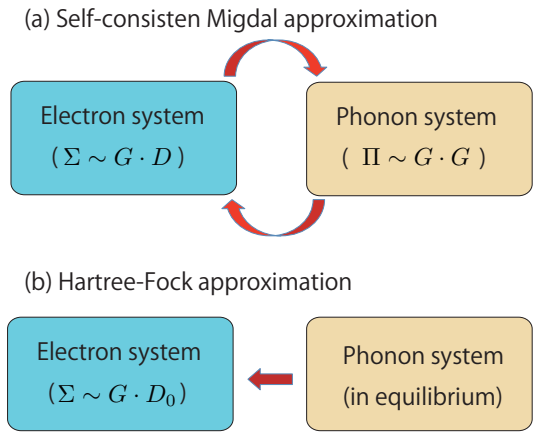}
  \caption{Schematic pictures of the
 self-consistent Migdal approximation (a) and the Hartree-Fock approximation (b). 
  $G$ denotes the dressed electron Green's function and $D$ the dressed phonon Green's function, while $D_0$ is the bare (equilibrium) phonon propagator. $\Sigma$ and $\Pi$ are the electron and phonon self-energies, respectively.}
  \label{fig:schema}
\end{figure}

An important point to note here is that the term ``Migdal approximation" is used for two distinct types of approximations in the literature on the Holstein model: One is the unrenormalized Migdal approximation,\cite{Freericks1993} where the {\it non-interacting} phonon propagator is used to express the effective interaction among the electrons. In other words, this approximation does {\it not} consider the phonon dynamics. This type of approximation has also been employed in recent studies of the dynamics of the Holstein model driven by strong laser fields.\cite{Sentef2013,Kemper2013,Kemper2014} 
The other is the {\it self-consistent} Migdal approximation,\cite{Freericks1994, Bauer2011, Hewson2002, Hewson2004, Capone2003,Hague2008} where the {\it dressed} phonon propagator is used and thus the phonon dynamics affects the electron self-energy and vice versa.  
In the following, we call the former approximation the Hartree-Fock (HF) approximation and the latter the Migdal approximation, see Fig.~\ref{fig:schema}. 
We will show that the Migdal approximation is more reliable than HF, by benchmarking equilibrium results against DMFT data obtained with a continuous-time quantum Monte Carlo (CT-QMC) impurity solver.\cite{Werner2007}
This is why we opt for the self-consistent Migdal approximation to discuss the dynamics of the isolated Holstein model after a sudden switch-on of the electron-phonon coupling. 

  %%%%%%%%%%%%%%%%%%%%%%%%%%%%%%%%%%%%%%%%%%
 %%%%%%%%%%%%%%%%%%%%%%%%%%%%%%%%%%%%%%%%%%

The main finding of this study is that {\it within the weak-coupling regime} there exist
a crossover between two different relaxation processes:
in the weaker electron-phonon coupling regime, the phonon oscillations are damped faster than the thermalization time of the electrons, which contrasts 
with the stronger interaction regime where the relaxation of the phonons becomes slower than the electron relaxation.   
We further demonstrate the importance of treating phonons dynamically by comparing the relaxation dynamics in the Migdal and HF approximations. It is shown that 
the phonon dynamics (i.e. the 
phonon self-energy) leads to qualitative changes in the relaxation dynamics. 

This paper is organized as follows. In Sec.~\ref{sec:model}, we introduce the Holstein model and discuss the nonequilibrium DMFT formalism along with the Migdal and HF impurity solvers for this model. 
%We also discuss several properties of the phonon propagator in the Kadanoff-Baym formalism.
 In Sec.~\ref{sec:result}, we first test the reliability of the Migdal and HF approximations in equilibrium. Then we explore the dynamics of the Holstein model after an interaction quench from the noninteracting state at zero temperature. 
We also show the difference between the Migdal and HF impurity solvers to discuss importance of phonon dynamics with their suitability for describing isolated electron-phonon systems.
 %We consider both the Migdal and HF impurity solvers to discuss their suitability for describing electron-phonon systems.
  Section~\ref{sec:conclude} provides a conclusion and outlook.

\section{\label{sec:model}Model and Formalism}
\subsection{Nonequilibrium DMFT for the Holstein model}
 
 The Hamiltonian for the Holstein model is 
 \begin{align}
 \mathcal{H}(t)=&-v\sum_{\langle  i,j\rangle,\sigma}(c_{i,\sigma}^{\dagger}c_{j,\sigma}+{\rm {\rm h.c.}})-\mu\sum_i n_i+\omega_0\sum_i a^{\dagger}_i a_i\nonumber\\
 &+g(t)\sum_i (a_i^{\dagger}+a_i)(n_{i,\uparrow}+n_{i,\downarrow}-\alpha),\label{eq:Holstein}
 \end{align}
where $c^{\dagger}_{i,\sigma}$ is the creation operator for an electron with spin $\sigma$ on site $i$, $v$ is the hopping parameter, $a^{\dagger}$ is the creation operator for a phonon with %the phonon 
frequency $\omega_0$, $g(t)$ is the (here time-dependent) electron-phonon interaction strength, and $\alpha$ is a constant 
that can be chosen arbitrarily, for which we take here 
$\alpha=\langle n_{\uparrow}+n_{\downarrow} \rangle$ so that the Hartree term in the self-energy vanishes. 
 We also note that in the anti-adiabatic limit ($\omega_0\rightarrow \infty$ with $\lambda\equiv2g^2/\omega_0$ fixed) the Holstein model becomes the attractive Hubbard model with a non-retarded interaction $-\lambda$. 
  It is also useful to introduce the position ($X$) and momentum ($P$) 
operators for the phonons, 
\begin{align}
X_i&=(a^{\dagger}_i+a_i)/\sqrt{2},\\
P_i&=i(a^{\dagger}_i-a_i)/\sqrt{2}.
\end{align}
%respectively. 

Throughout the paper, we assume the absence of long-range orders and focus on half-filling. 
We drive the system
out of equilibrium by changing the electron-phonon coupling constant $g$ from 
0 to a finite value $g_f$ at $t=0_+$. 
 To solve the problem, we employ the nonequilibrium DMFT.\cite{Aoki2013,Freericks2006} The DMFT formalism assumes a spatially local self-energy, and maps the lattice problem onto a quantum impurity model in a self-consistent manner. In order to describe the time evolution after a quench, one has to solve the DMFT equations on the L-shaped Kadanoff-Baym contour $\mathcal{C}$,
 which runs from $t=0$ up to the maximum simulation time $t_\text{max}$ along the real-time axis, back to $t=0$, and then proceeds to $-i\beta$ along the imaginary-time axis, where $\beta=1/T$ is the inverse temperature of the initial equilibrium state.  We define the electron Green's function $G_{i,j,\sigma}(t,t')$ and the local phonon Green's function $D(t,t')$ on this contour as 
 \begin{align}
 G_{i,j,\sigma}(t,t')&=-i\langle T_{\mathcal C} c_{i,\sigma}(t)c_{j,\sigma}^{\dagger}(t')\rangle,\\
 D(t,t')&=-2i\langle T_{\mathcal C} X_i(t)X_i(t')\rangle,
\label{D}
 \end{align}
where $T_{\mathcal C}$ is the contour-ordering operator.

The effective impurity action for the Holstein model is 
   \begin{align}
 S_{\rm{imp}}&=i\int_\mathcal{C} dt dt' \sum_{\sigma} d^{\dagger}_{\sigma}(t) \mathcal{G}^{-1}_{0,\sigma} (t,t') d_{\sigma}(t')\nonumber\\
& +i\int_\mathcal{C} dt a^{\dagger}(t)(i\partial_{t}-\omega_0) a(t)\nonumber\\
&-i\int_\mathcal{C} dt g(t)\left[a^{\dagger}(t)+a(t)\right]\left[n_{\uparrow}(t)+n_{\downarrow}(t)-\alpha\right],
  \end{align}
where the integrals run along the contour $\mathcal{C}$, $d_\sigma^\dagger$ is the creation operator for electrons at the impurity site,
and $\mathcal{G}_{0,\sigma}$ is the Weiss Green's function for the impurity problem, which is related to the hybridization function $\Delta_{\sigma}(t,t')$ by  
\begin{align}
\mathcal{G}^{-1}_{0,\sigma} (t,t') =(i\partial_{t}+\mu)\delta_\mathcal{C}(t,t')-\Delta_{\sigma}(t,t'),
\end{align}
where $\delta_\mathcal{C}$ is the delta-function on $\mathcal{C}$.
We can simplify the form of the action by expressing $a$ and $a^{\dagger}$ 
in terms of $X$ and $P$ and then tracing out $P$. This yields
 \begin{align}
S'_{\rm{imp}}=&i\int_\mathcal{C} dt dt' \sum_{\sigma} d^{\dagger}_{\sigma}(t) \mathcal{G}^{-1}_{0,\sigma} (t,t') d_{\sigma}(t')\nonumber\\
&+i\int_\mathcal{C} dt dt' X(t) D_0^{-1}(t,t')  X(t')\nonumber\\
&-i\sqrt{2}\int_\mathcal{C} dt g(t)X(t)\left[n_{\uparrow}(t)+n_{\downarrow}(t)-\alpha\right],\label{eq:imp_problem2}
\end{align}
where 
\begin{align}
D_0^{-1}(t,t')=\frac{-\partial^{2}_{t}-\omega^2_0}{2\omega_0}\delta_\mathcal{C}(t,t')
\end{align}
is the inverse of the bare phonon Green's function. 
In the solution of the DMFT equations, it is thus enough to consider the Green's function (\ref{D}) for the phonons.

Since the electrons interact with each other through the phonons and vice versa, to obtain the interacting Green's functions, we introduce the self-energies $\Sigma_{\sigma}(t,t')$ and $\Pi(t,t')$ for the electrons and phonons, respectively.
These functions satisfy the Dyson equations, 
\begin{align}
G_{\rm{imp},\sigma}(t,t')&=\mathcal{G}_{0,\sigma}(t,t')+[\mathcal{G}_{0,\sigma}*\Sigma_{\sigma} *G_{\rm{imp},\sigma}](t,t'),
\label{impurity Dyson}
\\
D(t,t')&=D_0(t,t')+[D_0*\Pi*D](t,t'),
\end{align}
where $*$ denotes the convolution on the contour $\mathcal{C}$. Here, the bare phonon Green's function can be expressed as 
\begin{align}
D_0(t,t')=&-i[\theta_\mathcal{C}(t,t')+f_B(\omega_0)]\exp\Big(-i\omega_0\int_{\mathcal{C},t'}^{t}dt_1\Big)\nonumber\\
&-i[\theta_\mathcal{C}(t',t)+f_B(\omega_0)]\exp\Big(-i\omega_0\int_{\mathcal{C},t}^{t'}dt_1\Big),
\end{align}
where $\theta_{\mathcal C}$ is the Heaviside function on $\mathcal{C}$, 
and $f_B(\omega_0)=(e^{\beta\omega_0}-1)^{-1}$ is the Bose distribution function.

The Weiss Green's function $\mathcal{G}_0$ (or hybridization function $\Delta$) is determined self-consistently in such a way that the electron Green's function for the impurity ($G_{{\rm imp}}$) becomes identical to the local electron Green's function of the lattice $G_{\rm{loc}}\equiv G_{i,i,\sigma}$, where the self-energy of the lattice system is identified with that of the effective impurity problem (self-consistent condition). 
 Here, we omit the site index in $G_{\rm{loc}}$ and also in $D$, assuming a homogeneous state.  
In this paper, we consider the Bethe lattice with infinite coordination number ($z\to\infty$), where the self-consistency condition simplifies to\cite{Georges1996,Aoki2013}
\begin{align}
\Delta_{\sigma}(t,t')=v_\ast^2 G_{\rm{loc},\sigma}(t,t')\label{eq:bethe}
\end{align}
with $v=v_\ast/\sqrt{z}$. In this case, the density of states is semi-elliptic, 
$\frac{1}{2\pi v^2_{\ast}} \sqrt{4v^2_\ast-\epsilon^2}$,
and we set $v_\ast=1$ in the following. Especially, the band-width $W$ is 4.

%%%%%%%%%%%%%%%%%%%%%%%%%%%%
\subsection{Observables}

  {\it Kinetic energy---} 
  By comparing the Dyson equations for the lattice and for the effective impurity problem, we obtain the expression for the electron kinetic energy, 
  \begin{align}
E_{{\rm kin}}(t)&\equiv\frac{-v}{N}\sum_{\langle i,j \rangle,\sigma}[\langle c_{i,\sigma}^{\dagger}(t)c_{j,\sigma}(t)\rangle+{\rm h.c.}]\nonumber\\
&=\frac{-i}{N}\sum_{i,\sigma}[\Delta_{i,\sigma}*G_{i,i,\sigma}]^{<}(t,t), \label{eq:ekin}
\end{align}
where $N$ is the 
number of lattice sites, and $<$ denotes lesser components.

{\it Electron-phonon correlation---}
From the equation of motion, $\partial_{t}c_{i,\sigma}(t)=i[\mathcal H(t),c_{i,\sigma}(t)]$,
we obtain
\begin{align}
&i\partial_{t}G_{i,i,\sigma}(t,t')|_{t'=t+0_{\mathcal C}^+}\nonumber\\
&=iv\sum_{j\text{ nn } i}  \langle c_{i,\sigma}^{\dagger}(t)c_{j,\sigma}(t)\rangle
-i\mu \langle c_{i,\sigma}^{\dagger}(t) c_{i,\sigma}(t)\rangle\nonumber\\
&\quad+i\sqrt{2}g(t)\langle X(t)c_{i,\sigma}^{\dagger}(t)c_{i,\sigma}(t)\rangle.
\label{equation of motion}
\end{align}
where $\sum_{j\text{ nn } i} $ is the summation over the nearest neighbors of site $i$. 
We now compare Eq.(\ref{equation of motion}) with the Dyson equation (\ref{impurity Dyson}) for the impurity problem and use the expression (\ref{eq:ekin}) for the kinetic energy to obtain 
\begin{align}
i\sqrt{2}g(t)\langle X(t)c_{i,\sigma}^{\dagger}(t)c_{i,\sigma}(t) \rangle=[\Sigma_{i,\sigma}*G_{i,i,\sigma}]^{<}(t,t).
\end{align}

{\it Phonon density---}
The density of phonons can be expressed in terms of the $X$ and $P$ as 
\begin{align}
\langle a^{\dagger} (t) a (t)\rangle=\frac{1}{2}[\langle X(t)X(t)\rangle+\langle P(t)P(t)\rangle]-\frac{1}{2},
\label{eq:phonon_dens}
\end{align}
where $\langle X(t)X(t)\rangle$ is obtained from $D(t,t)$, while $\langle P(t)P(t)\rangle$ is calculated  
from a second derivative of $D(t,t')$, as explained in Appendix C.

{\it Total energy---}
The total energy per site is given by
\begin{align}
E_\text{tot}(t)&=E_\text{kin}(t)-\mu\frac{1}{N}\sum_i\langle n_i \rangle+\frac{1}{N}\omega_0\sum_i \langle a^{\dagger}_i(t)a_i(t)\rangle \nonumber\\
&+\frac{\sqrt{2}}{N}\sum_{i}g(t)\left[\sum_{\sigma}\langle X_i(t) c_{i,\sigma}^{\dagger}(t)c_{i,\sigma}(t) \rangle-\alpha \langle X_i(t) \rangle\right].
\end{align}

 %%%%%%%%%%%%%%%%%%%%%%%%%%

\subsection{Impurity Solvers}\label{solvers}
\begin{figure}[t]
    \centering
   \includegraphics[width=60mm]{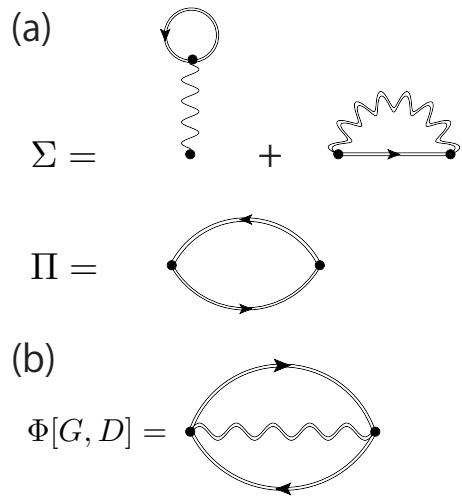}
  \caption{(a) The electron self-energy ($\Sigma$) 
and phonon self-energy ($\Pi$) diagrams in the self-consistent Migdal approximation. Here, the tadpole diagram should be evaluated as $-iG_{{\rm imp}}(t,t\!+\!0_\mathcal{C}^+)-\alpha/2$. (b) The Luttinger-Ward functional $\Phi$ for the self-consistent Migdal approximation. }
  \label{fig:ME_normal}
\end{figure}

The most demanding step in the DMFT self-consistency loop is the solution of the effective impurity problem, Eq. (\ref{eq:imp_problem2}). A numerically exact solution could in principle be obtained with the real-time CT-QMC method, as in the Hubbard model.\cite{Werner2009,Eckstein2009,Eckstein2010} 
However, QMC suffers from a sign (phase) 
problem when implemented on the real axis, which will make it very difficult to reach the relatively long times required to simulate phonon dynamics. Therefore, we employ here two different approximate diagrammatic impurity solvers (weak-coupling approximations):
 
{\it 1. (Self-consistent) Migdal approximation---}
The Feynman diagrams for the electron and phonon self-energies in the (self-consistent) Migdal approximation are shown in Fig.~\ref{fig:ME_normal}(a). The corresponding formulas read
 \begin{align}
 &\Sigma(t,t')=\nonumber\\
 &\!-\!\delta_\mathcal{C}(t,t')g(t) \!\!\int_\mathcal{C} \!dt_1 [\alpha+2iG_{{\rm imp}}(t_1,t_1\!+\!0_\mathcal{C}^+)]D_0(t_1,t)g(t_1)\nonumber\\
 &+iD(t,t')G_{{\rm imp}}(t,t')g(t)g(t'),\\
 \nonumber\\
& \Pi(t,t')=-i2g(t)g(t')G_{{\rm imp}}(t,t')G_{{\rm imp}}(t',t).
 \end{align}
This approximation has been used to study the Holstein model in equilibrium, and its accuracy has been discussed in a number of papers.\cite{Freericks1994, Bauer2011, Hewson2002, Hewson2004, Capone2003,Hague2008} As long as $g$ is not close to the critical value $g_c$ for the transition to the bipolaronic insulating phase and $\omega_0$ is small compared to the electron bandwidth, it provides a qualitatively good description.\cite{Hewson2004, Bauer2011} Since the self-energies of the electrons and phonons involve dressed propagators (as sketched in Fig.~\ref{fig:schema}(a)), we can take account of the interplay between the electrons and phonons in the dynamics.
With the choice of $\alpha=\langle n_{\uparrow}+n_{\downarrow} \rangle$  in Eq.~(\ref{eq:Holstein}) the Hartree term vanishes, and we can define a Luttinger-Ward  functional $\Phi[G,D]$ in this approximation as displayed in Fig.~\ref{fig:ME_normal}(b).
Hence the Migdal approximation is a conserving one.

%%%%%%%%%%%%%%%%%%%%%%%%%%%%%%%%%%%%%%%%%%
%%%%%%%%%%%%%%%%%%%%%%%%%%%%%%%%%%%%%%%%%%

 {\it 2. Hartree-Fock approximation---}
 As we mentioned in the introduction, the Hartree-Fock (HF) approximation is also sometimes called the (unrenormalized) Migdal approximation.\cite{Freericks1993} 
 In this approximation, the electron self-energy is given by
  \begin{align}
 &\Sigma(t,t')=\nonumber\\
 &\!-\!\delta_c(t,t')g(t)\!\!\int_c \!dt_1 [\alpha+2iG_{\rm{imp}}(t_1,t_1\!+\!0_c^+)]D_0(t_1,t)g(t_1)\nonumber\\
 &+iD_0(t,t')G_{\rm{imp}}(t,t')g(t)g(t').
 \end{align}
 The Feynman diagrams for the self-energy have the same structure as in Fig.~\ref{fig:ME_normal}(a), but the dressed phonon propagator is replaced with the bare equilibrium propagator.
Thus, in the HF approximation, we ignore the phonon self-energy, which 
means there is no feedback from the electrons to the phonons (Fig.~\ref{fig:schema}(b)). Hence we cannot extract 
the dynamics for the phonons from this scheme. Also, the HF approximation cannot be derived from a Luttinger-Ward functional, 
and is thus not conserving. 

The HF approximation has been used to study the equilibrium states\cite{Freericks1993} and nonequilibrium dynamics\cite{Kemper2014} of the Holstein model. 
In addition, the HF self-energy for small $g$ has  been added in some DMFT studies to describe the effect of a bosonic heat bath on the electrons.\cite{Eckstein2013, Eckstein2013b} 
The results in Section~\ref{HF} will confirm that the phonon effectively act as heat bath within the HF approximation.

\section{Results\label{sec:result}}
In what follows, we focus on the case of $\omega_0=0.7$ $(<W=4)$ and half-filling. 
We have checked that our discussion and the results are also applicable to smaller phonon frequencies such as $\omega_0=0.4$.  
We consider the weak-coupling regime, i.e. systems with coupling $g$ considerably smaller than the critical coupling $g_c$ for the bipolaronic transition (for which 
CT-QMC calculations give values $0.8\lesssim g_c \lesssim 0.85$  for 
$10<\beta<40$), but still with significant electron correlations. 
\subsection{Equilibrium}

  %%%%%%%%%%%%%%%%%%%%%%%%%%%%%%%%%%%%%%%%%%
 \begin{figure}[t]
   \centering
   \includegraphics[width=85mm]{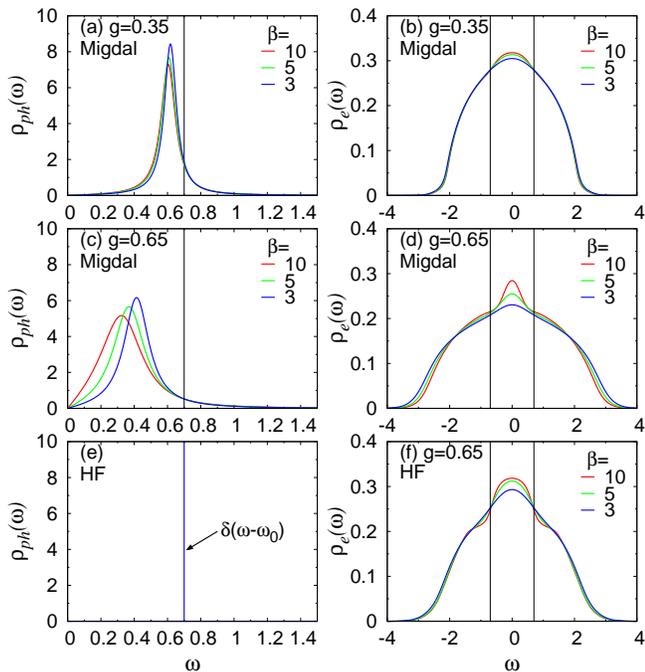}
  
  \caption{(a)(c) The phonon spectral functions $\rho_{\rm{ph}}(\omega)$ and (b)(d) the electron spectral functions $\rho_{e}(\omega)$ computed with the self-consistent Migdal approximation at half filling with $\omega_0=0.7$ 
for various values of $g$ and $T=1/\beta$. 
(e)(f) are corresponding results in the HF approximation at half filling with $g=0.65$. Vertical lines in each panel show the bare phonon frequency, $|\omega|=\omega_0$. }
  \label{fig:eq_spectrums_w0.7}
\end{figure}
%%%%%%%%%%%%%%%%%%%%%%%%%%%%%%%%%%%%%%%%%%

 %%%%%%%%%%%%%%%%%%%%%%%%%%%%%%%%%%%%%%%%%%
 \begin{figure}[t]
    \centering
   \includegraphics[width=85mm]{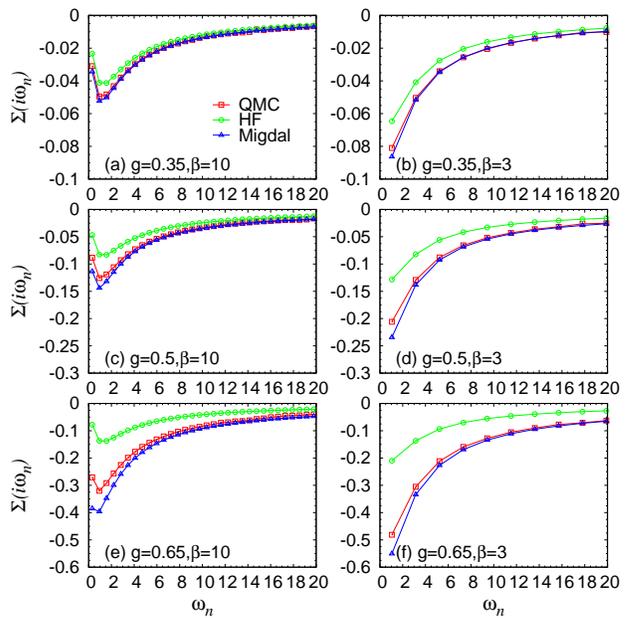}
 
  \caption{The electron self-energies on the Matsubara axis calculated in DMFT with the Migdal approximation, HF approximation, and CT-QMC impurity solvers for $\omega_0=0.7$ and indicated values of $g$ and $\beta$. 
  }
  \label{fig:Sig_comp_w0.7}
\end{figure}
%%%%%%%%%%%%%%%%%%%%%%%%%%%%%%%%%%%%%%%%%%

In this section, we benchmark the reliability of the Migdal and HF approximations as impurity solvers for DMFT,
and clarify which properties are correctly captured by these methods.
To this end, we consider equilibrium properties and first discuss the spectral functions computed with the two approximations. 
The spectral functions are defined by
\begin{align}
\rho_{\rm{ph}}(\omega)= -{\rm Im}D^R(\omega)/\pi,\\
\rho_\text{e}(\omega) = -{\rm Im} G^R_{{\rm loc}}(\omega)/\pi
\end{align} 
for phonons and electrons, respectively, and the superscript $R$ denotes retarded components. 
We obtain these spectral functions by calculating the equilibrium propagators on the real-time axis 
and performing Fourier transformations. 

In  Fig.~\ref{fig:eq_spectrums_w0.7} (a)(c), we plot
the phonon spectral functions $\rho_{\rm{ph}}(\omega)$ calculated with the Migdal approximation at half filling for $\omega_0=0.7$ and indicated values of $T=1/\beta$ and $g$. We find a single peak at a renormalized phonon frequency, 
which we call $\omega_0^r$, which shifts from $\omega=\omega_0$ (vertical lines) with increasing electron-phonon coupling $g$.  
This result is consistent with previous $T=0$ calculations based on the numerical renormalization group\cite{Hewson2002,Hewson2004} and the Migdal theory.\cite{Hague2008} As the temperature is increased, the phonon frequency becomes less renormalized, which indicates that the electron-phonon correlations are weaker at higher temperatures. The temperature dependence becomes more significant for larger $g$. 
In the HF approximation, by contrast, $\rho_{\rm ph}(\omega)$ has a delta-function peak at $\omega=\pm\omega_0$ [Fig.~\ref{fig:eq_spectrums_w0.7}(d)], since the phonons are assumed to have no self-energy.

The electron spectral functions $\rho_\text{e}(\omega)$ are shown in Fig.~\ref{fig:eq_spectrums_w0.7}(b)(d) for the Migdal approximation and in Fig.~\ref{fig:eq_spectrums_w0.7}(f) for the HF approximation. In both cases, there emerges a peak in the spectrum in the energy interval $|\omega|\lesssim \omega_0$ as the temperature is lowered. This peak represents quasiparticles (polarons) and
becomes more pronounced for stronger $g$. In the Migdal approximation, the peak becomes narrower with increasing $g$, which reflects the renormalization of the phonon frequency ($\omega_0^r$), while in the HF approximation the width is determined by the bare phonon frequency $\omega_0$.

In order to estimate the reliability of the Migdal and HF approximations, we show the corresponding self-energies on the Matsubara axis in Fig.~\ref{fig:Sig_comp_w0.7}. We compare them with the result computed using CT-QMC,\cite{Werner2007} which, being exact within statistical errors, serves as a reference.  
Clearly, the Migdal approximation is much closer to the CT-QMC results in the parameter regime considered here, 
where the HF approximation underestimates the electron self-energy, while the Migdal approximation slightly overestimates it. The quantitative difference becomes clearer as the interaction $g$ is increased. We conclude from these tests that the self-consistent Migdal approximation is more reliable and accurate than HF in a wide parameter regime when we are not too close to the bipolaronic phase boundary. 
We therefore expect that the Migdal approximation also provides a better description of the nonequilibrium dynamics than the HF approximation.

\subsection{Interaction quench: results of DMFT + Migdal approximation}

In this section, we study the time evolution of the Holstein model after an interaction quench of the electron-phonon coupling $g=0\to g_f$ at $t=0_+$
using the self-consistent Migdal approximation as an impurity solver.
The system is initially noninteracting and at equilibrium with temperature $T=0$. 
Although the length of the imaginary branch of the contour $\mathcal C$ is infinite in this case ($\beta=\infty$),
we can still treat the noninteracting initial state numerically, since 
the retarded ($R$) and lesser ($<$)  components are decoupled from the Matsubara (M) and left-mixing ($\rceil$) components in the Dyson equations
because
$\Sigma^M,\Sigma^{\rceil},\Pi^{M},\Pi^{\rceil}=0$. 
Similar quench problems have been studied for the Hubbard model with DMFT+QMC.\cite{Eckstein2009,Eckstein2010} With this initial condition, the momentum distribution $n(\epsilon_{\bm k},t)=-iG_{\bm k}^<(t,t)$ exhibits a discontinuous jump at  $\epsilon=0$ (i.e., the Fermi surface) for short times, while it is expected to become a smooth function once the system has thermalized at some nonzero temperature. The height of the jump $\Delta n(t)$ is thus a useful quantity that allows one to keep track of the thermalization process and to measure how much $n(\epsilon,t)$ deviates from the thermal distribution. In the following, we compare the relaxation of local observables to that of $n(\epsilon,t)$. 

\subsubsection{Local observables}
\begin{figure}[t]
    \centering
   \includegraphics[width=85mm]{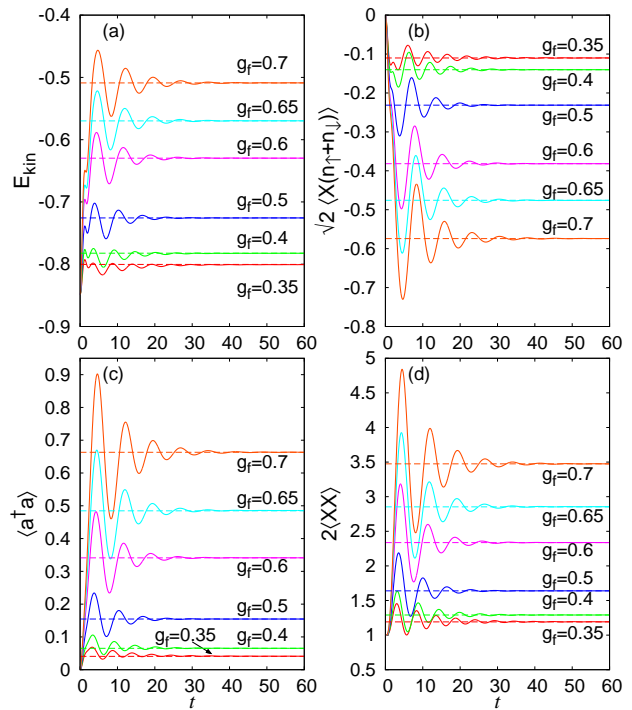}
 \caption{Temporal evolution of local quantities after an interaction quench from $g=0$ to indicated values of $g_f$ at $T=0$ with $\omega_0=0.7$: (a) $E_{\rm kin}$, (b) $\sqrt{2}\langle X(n_{\uparrow}+n_{\downarrow})\rangle$, (c) $\langle a^{\dagger}a \rangle$, and (d) $2\langle XX \rangle$. Dashed lines in each panel indicate the expected thermal values for each value of $g_f$.  }
  \label{fig:quantity_dynamics_w0.7}
\end{figure}

In Fig.~\ref{fig:quantity_dynamics_w0.7}, we show the time evolution of the kinetic energy, the correlation between the phonon displacement and the density of electrons $\sqrt{2}\langle X(n_{\uparrow}+n_{\downarrow})\rangle $, the phonon density $\langle a^{\dagger}a \rangle$, and 
the variance of the phonon displacement $2\langle XX\rangle$. 
All these local observables show coherent oscillations with twice the renormalized phonon frequency, $2\omega_0^r$. 
This can be explained as follows. 
First, let us suppose that each local phonon oscillates as $X(t)\sim \cos(\omega_0^r t)$.
Since the interaction quench does not discriminate the direction of the lattice distortion 
($X>0$ or $X<0$), the statistical distribution for the lattice displacement, $F(X,t)$, should be even in $X$, and oscillating with a 
period
$\pi/\omega_0^r$.
Due to the particle-hole symmetry  ($c\leftrightarrow c^{\dagger}, X\leftrightarrow -X$), this in fact exactly holds in our case and explains the oscillation of $2\langle XX\rangle$ with frequency $2\omega_0^r$. 
Provided that the phonon dynamics affects the electronic states through $F(X)$, 
it is natural to also expect oscillations of the other quantities with frequency $2\omega_0^r$.

 %%%%%%%%%%%%%%%%%%%%%%%%%%%%%%%%%%%%%%%%%%
%%%%%%%%%%%%%%%%%%%%%%%%%%%%%%%%%%%%%%%%%%  

 %%%%%%%%%%%%%%%%%%%%%%%%%%%%%%%%%%%%%%%%%%
 \begin{figure*}[t]
  \centering
   \includegraphics[width=160mm]{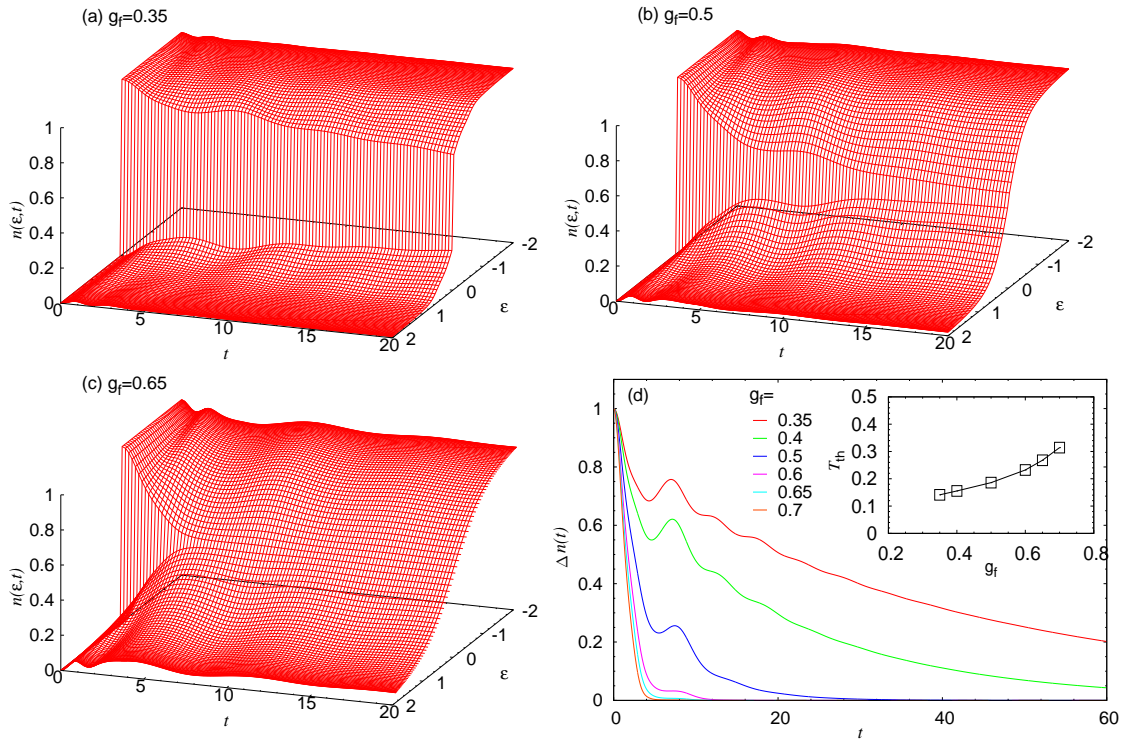}
  \caption{(a)-(c) Temporal evolution  of the momentum distribution for $\omega_0=0.7$ after quenches to various $g_f$. (d) Temporal evolution of the 
jump $\Delta n(t)$ for various values of $g_f$. The inset shows $T_{\rm th}$ against $g_f$.}
  \label{fig:T_0_w0.7_mom_dist_dynam}
\end{figure*}
%%%%%%%%%%%%%%%%%%%%%%%%%%%%%%%%%%%%%%%%%%
  
 As time evolves, the oscillations are damped and the amplitude of the oscillations becomes very small after $t=60$ in all the cases shown here. 
On the other hand, one expects that the system thermalizes, in the long-time limit, in an equilibrium state with temperature $T_{\rm th}$, which is defined by the relation 
 \begin{align}
E_{\rm tot}(t>0)=
\frac{{\rm Tr} e^{-\mathcal H_f/T_{\rm th}} \mathcal H_f}{{\rm Tr} e^{-\mathcal H_f/T_{\rm th}}},
\end{align}
 where ${\mathcal H_f}={\mathcal H(t>0)}$.
We note that the total energy is conserved after the quench, since the Hamiltonian is time-independent. The resultant $T_{\rm th}$, which increases with $g_f$, is shown in the inset of Fig.~\ref{fig:T_0_w0.7_mom_dist_dynam}(d). If the system thermalizes, expectation values of observables should approach those of the equilibrium state with $T_\text{th}$. 
For example, the thermal kinetic energy 
should approach 
\begin{align}
E_{\rm kin,th}=\frac{-v}{N}\sum_{\langle i,j\rangle,\sigma}
\frac{{\rm Tr} e^{-\mathcal H_f/T_{\rm th}} (c_{i,\sigma}^\dagger c_{j,\sigma}+\mbox{h.c.})}{{\rm Tr} e^{-\mathcal H_f/T_{\rm th}}}.
\label{E_kin,th}
\end{align}
The dashed lines in Fig.~\ref{fig:quantity_dynamics_w0.7} indicate the estimated thermal values at
$T_{\rm th}$ for each observable. It turns out that once the oscillations are well damped ($t\gtrsim 60$), the local observables are already very close to the thermal values.  However, we have to note that this does not necessarily mean that the system is fully thermalized, as we will discuss in the next section.

\subsubsection{Momentum distribution function}
In order to examine whether the system is really close to a thermalized state after the damping of the oscillations in the above local quantities,  
let us look at the evolution of the momentum distribution function for the electrons $n(\epsilon_{\bm k},t)=-iG_{\bm k}^<(t,t)$ [Fig.~\ref{fig:T_0_w0.7_mom_dist_dynam}(a)-(c)] and its jump $\Delta n(t)$ at $\epsilon=0$ [Fig.~\ref{fig:T_0_w0.7_mom_dist_dynam}(d)]. We start from $T=0$, $g=0$, so that $\Delta n(0)=1$. The jump does not immediately disappear after the quench, but decreases gradually. As we increase the interaction, $\Delta n(t)$ vanishes faster, as in the case of the Hubbard model.\cite{Eckstein2009,Eckstein2010} The main qualitative difference is that $\Delta n(t)$ oscillates in the present case of the Holstein model.

Now we are in a position to focus on the relation between the dynamics of $\Delta n(t)$ and that of the local observables. 
A key finding is that one can 
distinguish two qualitatively different types of relaxation behavior in the Holstein model {\it in the weak-coupling regime. }
The first type of relaxation dynamics is observed for couplings $g_f\lesssim 0.5$, where the long-time relaxation process is 
controlled by the electrons. 
At $g_f=0.35$, for instance, the phonon oscillation is damped and
the local (momentum integrated) quantities are essentially thermalized at $t=60$ (Fig.~\ref{fig:quantity_dynamics_w0.7}), while 
momentum-resolved observables for the electrons, exemplified by $\Delta n(t)$, are not thermalized [see Fig.~\ref{fig:T_0_w0.7_mom_dist_dynam} (d)]. In fact, $n(\epsilon,t)$ is still far from a thermal distribution at the longest accessible times.
While the height of the jump does not exhibit a plateau-like structure, as 
is the case in the Hubbard model,\cite{Eckstein2009,Eckstein2010} the observed behavior may still be regarded as a kind of prethermalization phenomenon in that 
local (momentum integrated) quantities thermalize fast, while momentum dependent quantities, such as $n(\epsilon,t)$, remain clearly nonthermal. 
These key observations characterize the relaxation behavior of the Holstein model in the sufficiently weak-coupling regime.

\begin{figure}[t]
\centering
   \includegraphics[width=85mm]{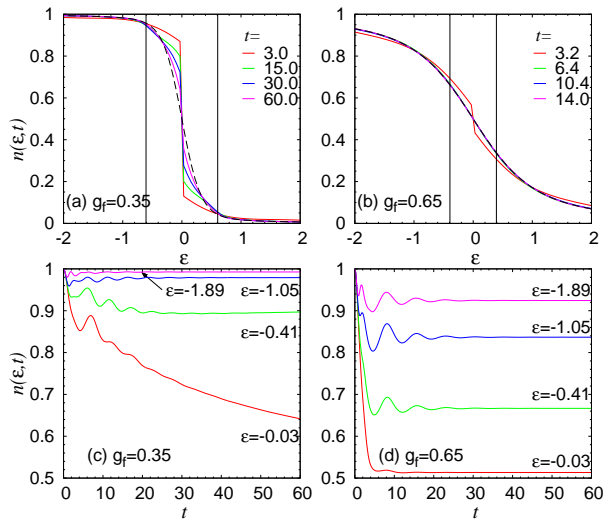}
  \caption{(a)(b) Electron momentum distribution function at various values of $t$ for $g_f=0.35$ (a) and $0.65$ (b). For (b) we chose special times, $\tilde{t}$, at which $E_{\rm kin}(\tilde t)=E_{\rm kin,th}$. The dashed lines in (a)(b) show the momentum distribution in equilibrium at $T_{\rm{th}}$ (which 
is invisible in (b) due to an almost perfect overlap with the data for $t\ge6.4$). Vertical lines show $|\epsilon|= \omega^r_0$.
(c)(d) Temporal evolution of $n(\epsilon,t)$ for several values of $\epsilon$ with $g_f=0.35$ (c) or $0.65$ (d).
}
  \label{fig:mom_dist_Mig_w0.7_detail}
\end{figure}

In Fig.~\ref{fig:mom_dist_Mig_w0.7_detail}(a)(c), 
we take a closer look at the relaxation of the momentum distribution $n(\epsilon,t)$ for $g_f=0.35$. 
Figure \ref{fig:mom_dist_Mig_w0.7_detail}(a) shows $n(\epsilon,t)$ for various values of $t$, 
while Fig.~\ref{fig:mom_dist_Mig_w0.7_detail}(c) shows the evolution of $n(\epsilon,t)$ for various values of $\epsilon$. 
The relaxation time strongly depends on the energy $\epsilon$: When $\epsilon \gtrsim \omega_0^{r}$ 
the electron relaxation is fast, while for $\epsilon \lesssim \omega_0^{r}$ the relaxation is slow, see Fig.~\ref{fig:mom_dist_Mig_w0.7_detail}(a). 
This behavior is similar to the HF results discussed in Ref.~\onlinecite{Kemper2013}.

%%%%%%%%%%%%%%%%%%%%%%%%%%%%%%%%%%%%%%%%%%
%%%%%%%%%%%%%%%%%%%%%%%%%%%%%%%%%%%%%%%%%%

A second type of relaxation behavior appears for stronger couplings ($g_f\gtrsim 0.5$, but still within the weak-coupling regime), where the phonons turn out to dominate the long-time dynamics.  
For example, at $g_f=0.65$, 
$\Delta n(t)$ vanishes [Fig.~\ref{fig:T_0_w0.7_mom_dist_dynam}(d)] before the oscillations of the momentum-integrated observables are damped (Fig.~\ref{fig:quantity_dynamics_w0.7}). Once these oscillations are fully damped, $n(\epsilon,t)$ also becomes equal to the thermal value. We have to note that the disappearance of the jump in $\Delta n(t)$ by no means implies that the distribution $n(\epsilon,t)$ is thermal. Rather, the distribution away from the Fermi energy continues to oscillate around its thermal value as shown in Fig.~\ref{fig:mom_dist_Mig_w0.7_detail}(d),
while $\Delta n(t)$ becomes very small before $t=8$ [Fig.~\ref{fig:T_0_w0.7_mom_dist_dynam}(d)].
The damping of the oscillations is related to the lifetime of phonons as will be discussed below in connection with the phonon self-energy. Hence the phonons, rather than electrons, 
govern the long-time relaxation in this regime. 
Interestingly, however, we can see in Fig. \ref{fig:mom_dist_Mig_w0.7_detail}(b) that 
$n(\epsilon,\tilde t)$ becomes almost indistinguishable from the thermalized distribution (dashed line, almost overlapping) at those times $\tilde t$ at which $E_{{\rm kin}}(\tilde t)=E_{{\rm kin, th}}$ [Eq.~(\ref{E_kin,th})] holds (after $\Delta n(t)$ has become negligible). 

The change from the electron-dominated to the phonon-dominated type of thermalization is a crossover,
i.e., the change is smooth and there is no abrupt change in the characteristics of the thermalization process, so that we can call the phenomenon 
a ``thermalization crossover".
In the present setup, the crossover %between these two types 
occurs around $g_f\sim 0.5$, 
where the oscillations and $\Delta n(t)$ vanish on similar time scales.

At this point we can comment on the relation between the present result 
and the phenomenological two-temperature model.\cite{Allen1987} We first note that our situation is rather different from what is assumed in the two-temperature model. In the latter, the assumption is that the electron degrees of freedom thermalize fast because of the Coulomb interaction, while in our case we only consider the electron-phonon coupling and no electron-electron interaction. 
Still, it is worthwhile to discuss the 
relation between the two models.
In the first type of relaxation ($g_f\lesssim 0.5$), in the Holstein model, the relaxation time strongly depends on $\epsilon$, and it would not be proper to describe it by a single decay rate as in the two-temperature model. 
More importantly, it is difficult to define a meaningful effective temperature for the electrons in this case
because of the $\epsilon$-dependent relaxation of $n(\epsilon,t)$.
In the second type of relaxation ($g_f\gtrsim 0.5$), the long-time behavior is dominated by damped oscillations. However, the two temperature model does not predict any oscillations, but rather a monotonic relaxation to the thermal value.\cite{Allen1987} Therefore, we conclude that neither of the two relaxation behaviors found here are captured by the conventional two-temperature model.

%%%%%%%%%%%%%%%%%%%%%%%%%%%%%%%%%%%%%%%%%%
 \begin{figure}[t]
 \centering
   \includegraphics[width=85mm]{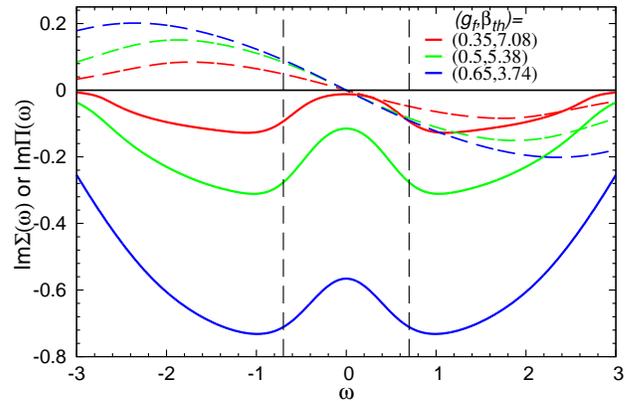}

  \caption{Imaginary parts of the electron self-energy (solid lines) and phonon self-energy (dashed lines)
  in equilibrium at $T_{\rm th}$ for various values of $g_f$. The vertical lines indicate $|\omega|=\omega_0$. }
  \label{fig:Mig_sig_rew_w0.7}
\end{figure}
%%%%%%%%%%%%%%%%%%%%%%%%%%%%%%%%%%%%%%%%%%

\subsubsection{Damping rates and self-energies}
In this section, we show that the different relaxation rates of physical quantities ($E_{\rm kin}, \Delta n(t)$...) can be related to the $g$-dependence of the electron ($\Sigma$) and phonon ($\Pi$) self-energies. 
In Fig.~\ref{fig:Mig_sig_rew_w0.7}, we plot the imaginary parts of  the electron self-energy and the phonon self-energy in equilibrium at $T=T_\text{th}$. 
When we look at the electron self-energy, we find that $\text{Im}\Sigma$ is relatively small in the energy range $|\omega|<|\omega_0^r|$, which becomes more evident at lower temperatures. This is consistent with the picture that electron (hole)-like quasiparticles cannot emit (absorb) phonons in this energy window, since the states below (above) the Fermi level are occupied (empty) at low enough temperatures. 
As a result, the quasiparticles survive for a long time in this energy range, since, roughly speaking, $-2\rm{Im}\Sigma(\omega=\epsilon)$ can be regarded as the relaxation rate.  We note that this picture qualitatively explains the different relaxation time scales of $n(\epsilon, t)$ for different $\epsilon$, illustrated in Fig.~\ref{fig:mom_dist_Mig_w0.7_detail}(a)(c).

  %%%%%%%%%%%%%%%%%%%%%%%%%%%%%%%%%%%%%%%%%%
 \begin{figure*}[t]
    \centering
   \includegraphics[width=160mm]{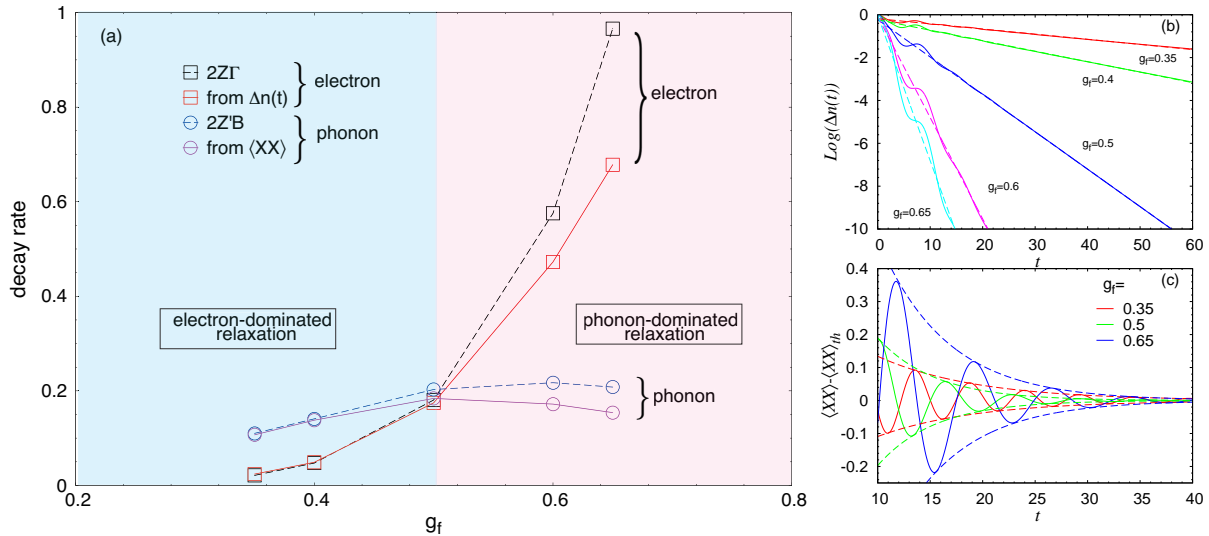}
\caption{(a) Electron and phonon energy scales (inverse relaxation times) against $g_f$.  (b) Temporal evolution of $\Delta n(t)$ on logarithmic scale 
for various values of $g_f$.  Dashed lines are exponential fits. 
(c) Temporal evolution of 
  $\langle XX\rangle-\langle XX \rangle_{\rm th}$.
  %The envelopes for the oscillation amplitude with exponential fits are shown with dashed curves.
  The dashed lines show exponential fits to the envelopes of the oscillating curves. 
  } 
  \label{fig:damp_ratio_Mig_w0.7}
\end{figure*}
%%%%%%%%%%%%%%%%%%%%%%%%%%%%%%%%%%%%%%%%%%
 
Now let us consider the lifetime of electron quasiparticles in more detail. If the quasiparticle picture is valid, the system, in the small $\omega$ regime, 
should have an electron self-energy of the form
 \begin{align}
 \Sigma^R(\omega)=(1-1/Z)\omega-i\Gamma+ O(\omega^2),
\end{align}
where $Z$ is the quasiparticle residue.
It follows that quasiparticles with momentum $\bm k$ have a renormalized energy $\epsilon_{\bm k}^r\equiv Z\epsilon_{\bm k}$ 
with a lifetime of $(2Z\Gamma )^{-1}$ at low-energies. 

The lifetime of phonons can be extracted 
from the phonon self-energy in a similar manner. 
For small enough $\omega$, the phonon self-energy can be expanded as 
\begin{align}
\Pi^R(\omega)=A-iB\frac{\omega}{\omega_0}+C\frac{\omega^2}{\omega^2_0}+O(\omega^3),
\end{align}
and the dressed Green's function becomes
\begin{align}
D^R(\omega)\simeq\frac{2Z'\omega_0}{(\omega-\omega'_0+i Z'B)(\omega+\omega'_0+i Z'B)+Z'^2 B^2},
\label{D^R}
\end{align}
where $\omega_0' = Z'\omega_0(1+A/\omega_0)^{1/2}$ 
and $Z'=(1-2C/\omega_0)^{-1}$.  Here, $\omega_0'$ is an approximation of the renormalized frequency $\omega^r_0$.
One can neglect the second term ($Z'^2B^2$) in the denominator of $D^R(\omega)$ (\ref{D^R})
if $B\ll\omega'_0$, 
since the absolute value of the first term in the denominator is at least 
$\sim O(B \omega'_0)$, which is much larger than the second term $\sim O(B^2)$.
It thus follows that $D^R(t)$ decays with a damping rate $Z'B$. We have checked this relation and found that $Z'B$ indeed explains the damping of $D^R(t)$ in the interaction regime considered. Hence the lifetime of phonons can be identified with $(2Z'B)^{-1}$.

Now we come to the key result of the present work.  
Figure~\ref{fig:damp_ratio_Mig_w0.7}(a) 
plots the quasiparticle lifetime for electrons ($2Z\Gamma$) and phonons ($2Z'B$), extracted from the equilibrium self-energies, against $g_f$.
Around $g_f=0.5$ the curves 
cross each other, so that 
$2Z\Gamma<2Z'B$ for $g_f<0.5$, while $2Z\Gamma>2Z'B$ for $g_f>0.5$. This means that the electrons (in the low-energy regime) decay 
more slowly than the phonons for $g_f<0.5$, while the phonons conversely decay 
more slowly than the electrons for $g_f>0.5$, as long as the quasiparticle picture is valid.

In Fig.~\ref{fig:damp_ratio_Mig_w0.7}(a), we also display the electron decay rate extracted from $\Delta n(t)$ by exponential fits, as shown in Fig.~\ref{fig:damp_ratio_Mig_w0.7}(b). Here we use the data from $t=0$ up to $t=60$ or up to 
$\Delta n(t)=10^{-4}$.  The decay rate increases with $g_f$, and 
matches the value of $2Z\Gamma$ to a good approximation. In the smaller-$g_f$ regime the discrepancy is very small, while $2Z\Gamma$ tends to overestimate the exponent of $\Delta n(t)$ in the larger-$g_f$ regime.  
On the other hand, the phonon decay rate $2Z'B$ is reflected in the damping of $E_{\rm kin}, \langle Xn\rangle, \langle XX \rangle$ and $\langle a^{\dagger}a\rangle$.
As an example, Fig.~\ref{fig:damp_ratio_Mig_w0.7}(c) displays the oscillations of $\langle XX \rangle-\langle XX \rangle_{\rm th}$.  
We fitted the envelopes with exponentials, and plot the corresponding decay rates in Fig.~\ref{fig:damp_ratio_Mig_w0.7}(a). 
The oscillations for other quantities ($E_{\rm kin}, \langle Xn\rangle$ and $\langle a^{\dagger}a\rangle$) have almost the same damping rates.  
As can be seen in  Fig.~\ref{fig:damp_ratio_Mig_w0.7}(a), $2Z'B$ indeed provides a good explanation for the damping rates of local quantities. To be more precise, while the agreement with $2Z'B$ is very good for  
$g_f\lesssim0.5$, 
the quasiparticle lifetime from the phonon self-energy tends to overestimate the damping rate of the $\langle XX \rangle-\langle XX \rangle_{\rm th}$ oscillations and this tendency becomes clearer as we increase $g_f$. 

The above analysis indicates that 
the different dependence of the electron and phonon lifetimes on the electron-phonon coupling $g$ explains 
the two different relaxation regimes: In the weaker-coupling regime, the lifetime for phonons is shorter than that for electrons, so that the phonon oscillations are damped before electron's $n(\epsilon,t)$ thermalizes (electron-dominated thermalization). 
In the stronger-coupling regime, the situation is reversed and 
the electron lifetime is shorter than the phonon lifetime (Fig.~\ref{fig:damp_ratio_Mig_w0.7}(a)). Hence $\Delta n(t)$ vanishes quickly and the momentum distribution approaches to its thermal value quickly. However, since the phonons are still in the process of relaxing and oscillating, the electrons are forced to move with them (phonon-dominated thermalization).

\subsubsection{Nonequilibrium spectral functions}
%%%%%%%%%%%%%%%%%%%%%%%%%%%%%%%%%%%%%%%%%%
 \begin{figure}[t]
  \centering
   \includegraphics[width=85mm]{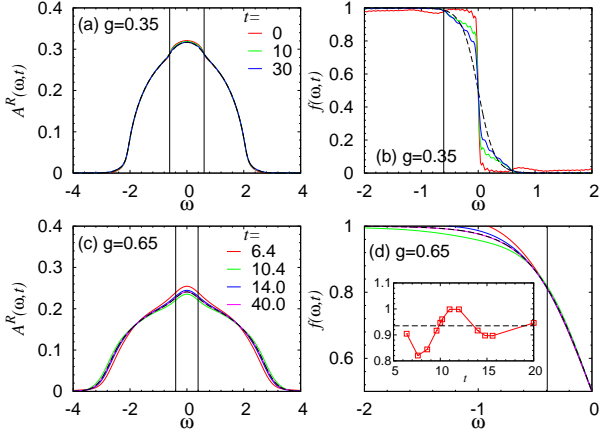}
  \caption{Nonequilibrium spectral function $A^{R}(\omega,t)$ at different $t$ for $g_f=0.35$ (a) or $0.65$ (c), and nonequilibrium distribution function 
$f(\omega,t)$ for $g_f=0.35$ (b) or $0.65$ (d). The dashed curves represent the thermal $A^R$ and distribution functions.
  Vertical lines in each panel indicate $\omega_0^r$ in equilibrium at $T_{\rm th}$ for each value of $g_f$. The inset in (d) plots the time evolution of  $\frac{\partial f(\omega, t)}{\partial \omega}|_{\omega=0}$, where the dotted line shows the thermal value.}
  \label{fig:Mig_T_0_spectrum}
\end{figure}
%%%%%%%%%%%%%%%%%%%%%%%%%%%%%%%%%%%%%%%%%%

We next discuss how the two different types of relaxation manifest themselves in the spectral function and nonequilibrium distribution function.  In nonequilibrium, we define the electron spectral function $A^R$ and occupied spectral function $A^<$ as 
\begin{align}
A^{R,<}(\omega,t)=\mp\frac{1}{\pi}{\rm Im}\int^{\infty}_t dt' e^{i\omega(t'-t)}G_{\rm loc}^{R,<}(t',t),
\end{align}
where $-$ is for $R$ and $+$ is for $<$. For a slowly varying state, $A^<(\omega,t)$ corresponds to the time-resolved photoemission spectrum and $A^R$ to the time-resolved total spectral function. 
From these one can define the ``nonequilibrium distribution function" $f(\omega, t)\equiv A^{<}(\omega,t)/A^{R}(\omega,t)$. 
In Fig.~\ref{fig:Mig_T_0_spectrum}, we display $A^{R}(\omega,t)$ and $f(\omega, t)$ at different times.  
The result for $g_f=0.35$ in the electron-dominated regime is shown in 
Fig.~\ref{fig:Mig_T_0_spectrum}(a,b). We first note that even at $t=0$, $A^{R}$ shows a peak structure around $\omega=0$ and therefore is different from the spectral function of the free system. This is because $A^{R}(\omega,t)$ includes information on later times than $t$. $A^{R}(\omega,t)$ and $f(\omega,t)$ relaxes to its thermal value quickly for $\omega\gtrsim\omega_0^r$, while for $\omega\lesssim\omega_0^r$ the relaxation is slow and gradual. This is consistent with the behavior of the momentum distribution and with a previous analysis of the photoexcited Holstein model.\cite{Kemper2014}  The small wiggles in Fig.~\ref{fig:Mig_T_0_spectrum}  
(b) are Fourier cutoff artifacts.

The dynamics for a larger $g_f=0.65$ in the phonon-dominated regime is  
shown in Fig.~\ref{fig:Mig_T_0_spectrum}(c,d). Here, we again choose the special times at which $E_{{\rm kin}}=E_{{\rm kin, th}}$.
Both $A^R(\omega, t)$ and $f(\omega, t)$ turn out to be different from the thermal curves, though the momentum distributions are indistinguishable from the thermal ones at these times (Fig. \ref{fig:mom_dist_Mig_w0.7_detail}(b)). This difference is not too surprising, since $A^{<}$ and $A^{R}$ are not determined by instantaneous temporal information unlike the momentum distribution function. We also show $\frac{\partial f(\omega, t)}{\partial \omega}|_{\omega=0}$ in the inset of the panel (d). The oscillation of this slope indicates that $f(\omega, t)$ near $\omega=0$ also oscillates around its thermal value. Therefore, in contrast to the relaxation in the weaker-coupling regime, $A^{R}(\omega,t)$ and $f(\omega,t)$ oscillate around their thermal values not only in the energy range $|\omega|\gtrsim\omega_0^r$ but also for $|\omega|\lesssim\omega_0^r$. Once this oscillation is fully damped (see $t=40$), they relax to the thermal $A^{R}$ and Fermi distribution function, respectively.

%%%%%%%%%%%%%%%%%%%%%%%
\subsection{Interaction quench: results of DMFT + Hartree-Fock approximation}
\label{HF}
%%%%%%%%%%%%%%%%%%%%%%%%%%%%%%%%%%%%%%%%%%
 \begin{figure}[t]
  \centering
   \includegraphics[width=85mm]{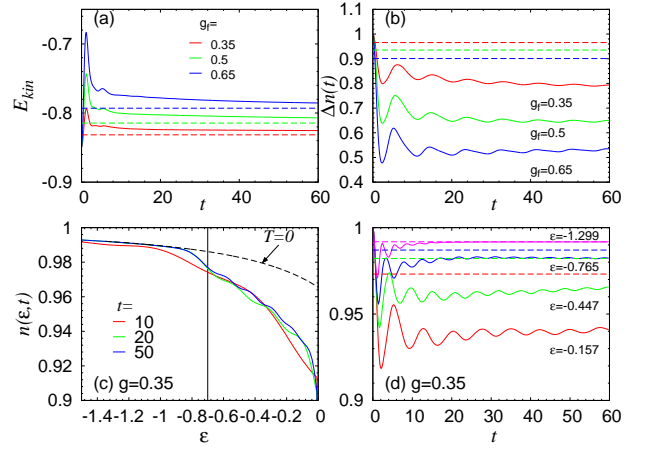}

  \caption{DMFT + HF result for the 
temporal evolution of (a) the kinetic energy, (b) $\Delta n(t)$, (c) $n(\epsilon,t)$ with $t$ fixed, and (d) $n(\epsilon,t)$ with $\epsilon$ fixed, after quenches to the indicated values of $g$ with $\omega_0=0.7$. The vertical line in panel (c) indicates $|\epsilon|=\omega_0$, while dashed curves in panels (c,d) indicate the thermal values at $T=0$.}
  \label{fig:mom_dist_HF_w0.7_detail}
\end{figure}
%%%%%%%%%%%%%%%%%%%%%%%%%%%%%%%%%%%%%%%%%%
In order to understand the effect of the phonon dynamics, 
let us compare the above results from the self-consistent 
Migdal scheme with those from the HF approximation.
In the latter, the phonons are treated as noninteracting equilibrium phonons, as discussed in Sec.~\ref{solvers}.
First of all, we note that, since the phonon is assumed to stay in equilibrium in HF, the thermalization crossover that we have revealed with the Migdal approximation does not occur.
In Fig.~\ref{fig:mom_dist_HF_w0.7_detail} (a), we show the HF result for 
the kinetic energy for $\omega_0=0.7$ and several values of $g_f$.
One striking difference from the self-consistent Migdal results [Fig.~\ref{fig:quantity_dynamics_w0.7}(a)] is that the oscillations are damped very quickly within $t<10$.
After that, 
the kinetic energy seems to slowly approach a steady value in the long-time limit. 
As discussed in Ref.~\onlinecite{Eckstein2013}, the HF self-energy is expected to act as a heat bath, which cools electrons down to the temperature of the initial equilibrium phonons (i.e., $T=0$ here). The results are indeed consistent with the system approaching the $T=0$ state.
In Fig.~\ref{fig:mom_dist_HF_w0.7_detail}(a), we plot the equilibrium values at $T=0$ by dashed lines, and it appears that both $E_{\rm kin}$ and $\langle X (n_{\uparrow}+n_{\downarrow})\rangle$ (not shown) gradually relax to the thermal values at $T=0$.

The HF results for $\Delta n(t)$ are shown in Fig.~\ref{fig:mom_dist_HF_w0.7_detail}(b).
After the quench, $\Delta n(t)$ starts to decrease, but remains large compared to the Migdal results [Fig.~\ref{fig:T_0_w0.7_mom_dist_dynam}(d)].
The fact that $\Delta n(t)$ does not vanish is consistent with the expectation that the phonons in HF effectively act as a heat bath with $T=0$.
On the other hand, $\Delta n(t)$ is still far from the expected thermal value for $T=0$ even at $t=60$, showing that the cooling rate is very low. This can be understood as follows.  
In the $T=0$ equilibrium state, we have ${\rm Im}\Sigma(\omega)\propto \omega^2 $ for the Fermi liquid. Hence, the decay rate for $\Delta n(t)$ is expected to become zero as the system approaches the equilibrium state at $T=0$.

Figure \ref{fig:mom_dist_HF_w0.7_detail}(c,d) shows $n(\epsilon,t)$ for $g_f=0.35$. The distribution at $|\epsilon|\gtrsim\omega_0$ relaxes faster to the equilibrium value at $T=0$
than that at $|\epsilon|\lesssim\omega_0$. This is similar to the Migdal result (Fig.~\ref{fig:mom_dist_Mig_w0.7_detail}) and the previous study (Ref.~\onlinecite{Sentef2013}). 
In addition, one finds in Fig.~\ref{fig:mom_dist_HF_w0.7_detail}(d) that $n(\epsilon,t)$ shows more pronounced oscillations than in the Migdal approximation, and that the 
$n(\epsilon,t)$ at different values of $\epsilon$ exhibit oscillations with different frequencies.\cite{Sentef2013}
The dephasing of the oscillations in momentum space leads to a complicated structure in $n(\epsilon,t)$ [Fig.~\ref{fig:mom_dist_HF_w0.7_detail}(c)] 
and a fast damping of the oscillations in $E_{\rm kin}$ [Fig.~\ref{fig:mom_dist_HF_w0.7_detail}(a)], which is a scenario different from the relaxation mechanism in the Migdal approximation.

The comparison of the HF and Migdal results implies that the feedback of the nonequilibrium phonons to the electrons leads to a qualitatively very different dynamics, so that the HF approximation cannot properly describe the evolution of isolated systems. However, we note that it may be possible to use the HF approximation as a phenomenological treatment for electrons coupled to a heat bath (open system), as in Refs.~\onlinecite{Eckstein2013, Eckstein2013b}.

\subsection{Discussions}
We have discussed the relaxation dynamics of the infinite-dimensional Holstein model based on the nonequilibrium DMFT.  
The DMFT analysis is limited because the method neglects the momentum dependence of the electron and phonon self-energies, an approximation which is justified in the limit of infinite spatial dimensions. When we consider a finite dimensional system, however, we need to take into account the momentum dependence. It turns out that the phonon self-energy can have a significant momentum dependence, as is clear from an evaluation of the lowest order phonon self-energy. In particular, the phonon self-energy vanishes in the zero wavelength limit.
Therefore, in order to discuss whether phonons or electrons are the bottleneck in the relaxation process in finite dimensions, we have to consider the momentum dependence of the self-energies. 
Nevertheless, our results can be expected to be applicable to systems in high dimensions or with large coordination numbers, where the momentum dependence 
is expected to be not so essential. Corrections from short-range spatial correlations can be captured by extending the nonequilibrium DMFT formalism to a nonequilibrium dynamical cluster approximation (DCA).\cite{Tsuji2014} This route will in principle provide a systematic way to include the momentum dependence.

Another limitation in our analysis is the use of the self-consistent Migdal approximation as an impurity solver for the nonequilibrium DMFT.
We have analyzed the $\omega_0$ dependence of the relaxation dynamics and quasiparticle lifetimes (not shown). It turns out that the thermalization crossover point moves to a smaller $\lambda(=\frac{2g^2}{\omega_0})$ as we decrease the phonon frequency $\omega_0$. In this regime, the Migdal approximation becomes quantitatively more reliable, so that our observation suggests that
the essential nature of the phenomenon (thermalization crossover) can be correctly described by the Migdal approximation. The reason why we choose relatively large $g$(or $\lambda$) and $\omega_0$ in this paper is because of computational limitations (limited accessible timescales), though we note that these parameters are still significantly smaller than the electron bandwidth and in the weak-coupling regime.

\section{Conclusions\label{sec:conclude}}
In this paper, we have studied the dynamics of the Holstein model after a quench (sudden switch-on) of the interaction using the nonequilibrium DMFT in the 
weak-coupling regime. As an impurity solver, we have employed the self-consistent Migdal approximation, which includes the dynamics of phonons via the phonon self-energy. It turns out that the local (momentum-summed) quantities exhibit essentially $2\omega_0^r$ oscillations (with $\omega_0^r$ being the renormalized phonon frequency).  
A key finding here is that there exists a thermalization crossover between two distinct 
regions as we vary the quenched electron-phonon coupling $g$ within the weak-coupling region: The smaller-$g$ region shows a fast damping of the oscillations originating from the phonon dynamics, with the momentum-summed quantities approaching the thermal values quickly, while the momentum distribution of the electrons exhibits a much slower relaxation (electron-dominated relaxation). The second region corresponds to larger $g$, but still in the weak-coupling regime (well before the phase transition to the bipolaronic phase). There, the jump in the momentum distribution quickly vanishes, and the momentum distribution quickly approaches its thermal value.  Since the phonon oscillations damp more slowly, the momentum distribution oscillates with the phonons around the thermal value (phonon-dominated relaxation).
We have revealed that the change in the relaxation behavior originates 
from a different $g$-dependence of the electron and phonon self-energies.  
We have further confirmed the importance of the phonon dynamics by comparing the self-consistent Migdal results with the HF results which do not include the phonon dynamics.
It turns out that the latter approximation describes a totally different type of relaxation process with phonons effectively acting as a heat bath.

Our work can serve as a benchmark for further studies of electron-phonon systems. The effect of additional terms such as the Coulomb interaction and acoustic phonons in the relaxation process will be important to understand. In addition, it will be interesting to study the dynamics of ordered phases in electron-phonon systems.\cite{Murakami2013,Murakami2014} This topic will be discussed in a separate publication.  

\acknowledgements
The authors would like to thank D. Golez and A. Millis for fruitful discussions.
YM, NT, and HA wish to thank the Fribourg University for hospitality when the manuscript was written. YM, NT and HA have been supported by
LEMSUPER (EU-Japan Superconductor Project) from JST, and 
YM is supported by JSPS Research Fellowships for Young Scientists and Advanced Leading Graduate Course for Photon Science (ALPS).
NT was supported by Grant-in-Aid for Scientific Research (Grant Nos. 25104709, 25800192).
PW acknowledges support from FP7/ERC starting Grant No. 278023.

\section{Appendix}
\subsection{Dyson equation}
To investigate the dynamics of a certain type of Green's function $\Theta(t,t')$, one needs to solve the Dyson equation, which can be expressed in the form  
\begin{align}
[1-F]*\Theta=Q.
\end{align}
If we explicitly write down this equation for the retarded (R), lesser ($<$) and left-mixing ($\rceil$) components, it becomes 
\begin{align}
&\Theta^{R}(t,t')-\int^{t}_{t'}d\bar{t}F^{R}(t,\bar{t})\Theta^{R}(\bar{t},t')=Q^{R}(t,t')\\
&\Theta^{<}(t,t')-\int^{t}_{0}d\bar{t}F^{R}(t,\bar{t})\Theta^{<}(\bar{t},t')=Q^{<}(t,t')\\
&+\int^{t'}_{0}d\bar{t}F^{<}(t,\bar{t})\Theta^{A}(\bar{t},t')-i\int^{\beta}_{0}d\bar{\tau}F^{\rceil}(t,\bar{\tau})\Theta^{\lceil}(\bar{\tau},t')\nonumber\\
&\Theta^{\rceil}(t,\tau')-\int^{t}_{0}d\bar{t}F^{R}(t,\bar{t})\Theta^{\rceil}(\bar{t},\tau')=Q^{\rceil}(t,\tau')\nonumber\\
&+\int^{\beta}_{0}d\bar{\tau}F^{\rceil}(t,\bar{\tau})\Theta^{M}(\bar{\tau},\tau').
\end{align}
When $\Theta$ corresponds to $G$ or $D$, the components $\Theta^A$ and $\Theta^{\lceil}$ are related to $\Theta^R$ and $\Theta^{\rceil}$ (although the relation is different for the fermionic and bosonic Green's function), 
so the above set of equations is closed, and we only need to solve these three equations, see Ref. \onlinecite{Aoki2013} and Appendix B.
%%%%%%%%%%%%%%
 \subsection{Properties of the phonon Green's function}
Here we explicitly state several relations of the phonon Green's function $D(t,t')$, which are important in the implementation of the Dyson equation for the phonon propagator. 
From Eq.~(\ref{D}), it follows that
 \begin{align}
D(t,t')=D(t',t),\label{eq:M4}
\end{align}
and therefore 
\begin{align}
D^{M}(\tau,\tau')&=D^{M}(\tau',\tau),\\
D^{A}(t',t)&=D^{R}(t,t'),\label{eq:ph_rel1}\\
D^{\lceil}(\tau',t)&=D^{\rceil}(t,\tau')\label{eq:ph_rel2}.
\end{align}
We also note that 
\begin{align}
D^{<}(t,t')^*&=-D^{<}(t',t).
\end{align}
Furthermore, in contrast to $G$, the retarded part of $D$ has no jump at $t=t'$, i.e., 
  \begin{align}
 D^{R}(t+0^+,t)&=D^{R}(t,t+0^+)=0.
 \end{align}
 
%%%%%%%%%%%%%%%%%%%%%%
\subsection{Derivatives of the phonon propagator}

Here we discuss the properties of the derivative of the phonon Green's function. With Eq.~(\ref{eq:Holstein}),
\begin{align}
i\partial_{t}X(t)=\omega_0(-a^{\dagger}(t)+a({t}))/\sqrt{2},
\end{align}
so we find 
\begin{align}
D_{\text{d1}}(t,t')&\equiv\frac{\partial_{t}D(t,t')}{\omega_0}\nonumber\\
&=-i2\langle T_{c} P(t)X(t')\rangle,
\end{align}
where $P=\frac{-a^{\dagger}+a}{i\sqrt{2}}$ and $[X, P]=i$.
In the same manner,
\begin{align}
D_{\text{d2}}(t,t')&\equiv\frac{\partial_{t'}D(t,t')}{\omega_0}\nonumber\\
&=-i2\langle T_{c} X(t)P(t')\rangle.\nonumber
\end{align}
In addition, we calculate the second derivative of $D$,
\begin{align}
D_{\text{d1,d2}}(t,t')&\equiv\frac{\partial_{t}\partial_{t'}D(t,t')}{\omega_0^2}\nonumber\\
&=\frac{2}{\omega_0}\delta_c(t,t')-i2\langle T_c P(t)P(t')\rangle.
\end{align}
If we know $D, D_{\rm{d1}}, D_{\rm{d2}}$ and $D_{\rm{d1,d2}}$ we can recover the usual boson Green's function defined as  $-i\langle T_c a(t)a^{\dagger}(t')\rangle$, and, in particular, the phonon density $\langle a^{\dagger}(t)a(t)\rangle$, Eq.~(\ref{eq:phonon_dens}).
These quantities can be evaluated with $D$ and $\Pi$ by considering the following equations,
\begin{align}
D_{d1}(t,t')&=D_{0,d1}(t,t')+[D_{0,d1}*\Pi*D](t,t'),\label{eq:d1}\\
D_{d2}(t,t')&=D_{0,d2}(t,t')+[D*\Pi*D_{0,d2}](t,t'),\label{eq:d2}\\
D_{d1,d2}(t,t')&=D_{0,d1,d2}(t,t')+[D_{0,d1}*\Pi*D_{d2}](t,t').\label{eq:d1d2}
\end{align}
Since
\begin{align}
D_{d1}^R(t,t')&=D_{d2}^A(t',t),\\
D_{d1}^<(t,t')&=-D_{d2}^<(t',t)^*,\\
D_{d1}^{\rceil}(t,\tau')&=D_{d2}^{\lceil}(\tau',t),
\end{align}
we can focus on the $R$, $<$ and $\rceil$ components in Eqs.~(\ref{eq:d1}), (\ref{eq:d2}) and (\ref{eq:d1d2}). 
In particular, to evaluate the phonon density $\langle a^{\dagger}a\rangle$, we only need to know $D_{d1,d2}^<$. In this case we only need to evaluate Eq.~(\ref{eq:d1}) and solve the $<$ part of Eq.~(\ref{eq:d1d2}).

\end{document}